# Manipulating the anisotropic phase separation in strained VO$_2$ epitaxial films by nanoscale ion-implantation

Changlong Hu[1+], Liang Li[1+], Xiaolei Wen[2], Yuliang Chen[1], Bowen Li[1], Hui Ren[1], Shanguang Zhao[1], Chongwen Zou[1]*

[1]*National Synchrotron Radiation Laboratory, University of Science and Technology of China, Hefei, 230029, China*

[2] *Center for Micro- and Nanoscale Research and Fabrication, University of Science and Technology of China*

Corresponding Author: czou@ustc.edu.cn
+These two authors contributed equally to this paper.

# ABSTRCT

Manipulating the strain induced poly-domains and phase transition in correlated oxide material are important for high performance devices fabrication. Though the electronic transport in the strained oxide film at macroscopic scales can be directly measured, the anisotropic electronic state and the controllable phase separation cross the insulator-to-metal transition within nanoscale size are still elusive. Here, we selected $VO_2$ crystal film as a prototypical oxide and achieved the manipulation of anisotropy electronic phase separation via injecting $He^+$ nanobeam into $VO_2$ film at room temperature. In addition, this nanoscale phase separation was directly visualized by infrared near-field imaging measurements, showing the pronounced and unique $c_R$-axis dependent anisotropy on $VO_2$ surface. Our results offered new insights towards understanding the anisotropic nanoscale phase separation in strained metal oxide films.

Correlated oxides have attracted tremendous attention due to the distinct emergent phenomena including the colossal magnetoresistance, insulator-metal transition and high-temperature superconductivity [1], which are closely associated with the degrees of freedom including the spin-charge-lattice-orbital interactions[2]. Specifically, the lattice strains are of particular tool owing to the characteristic that it can cause conspicuous new physics in materials. For example, strain can lower the correlated oxides transition temperature[3], the increase of the superconducting critical temperature [4], the enhancement of the ferroelectric polarization [5] or even producing giant pseudomagnetic fields [6]. Spontaneous formation of multi-domains by external compressive strain can produce many new macroscopic orders [7], leading to the coexistence of multiple constituent phases or pronounced phase separations in correlated materials [8].

As a typical correlated transition metal oxide, vanadium dioxide ($VO_2$), has attracted tremendous interest due to its special metal-insulator transition (MIT) near the room temperature (~340K) [9]. During the MIT process, $VO_2$ crystal will change from a low temperature monoclinic (*M*) structure to high temperature rutile (*R*) structure and the resistance will show a sharp drop with the magnitude of up to five orders [10]. It is also revealed that the $V^{4+}$ ions dimers will tilt with respect to the tetragonal $c_R$ axis across the MIT, causing the specimen to expand by 1% along the $c_R$ axis [11]. Thus the phase transition of $VO_2$ crystal exhibits pronounced responses to strain stimuli, which always induces multiple phases domains or pronounced phase seperations[12,33]. It is known that manipulate the macroscopic order or control the phase separations is technologically important for $VO_2$ device fabrication. While till now almost all of the previous studies are focusing on the observations of phase separations triggered by fixed interfacial strain in epitaxial film [12, 16, 17], the artificial manipulation of strain to control the phase transition in $VO_2$ crystal is rarely reported.

In the current study, we prepared $VO_2$ crystal film by an rf-plasma assisted oxide molecule beam epitaxy (OMBE) on $TiO_2$ (110) and (001) crystal substrates, respectively. By injecting $He^+$ nanobeam into $VO_2$ film, the local strain distribution around the injection zone can be manipulated, which leads to anisotropic phase separations. Moreover, these naonscale anisotropic phase separations can be directly observed by scattering-type scanning near-field optical microscope (s-SNOM) and the anisotropic features strongly dependent on the growth orientation. Combined with the strain simulation results, it is suggested that this anisotropic phase separation mainly due to the local strain field modulated by $He^+$ nanobeam injection, which is closely associated with the in-plane $c_R$ axis of $VO_2$. Our current studies not only reveal the relationship between the anisotropic phase separation and the $c_R$ axis of $VO_2$, but also demonstrate a strain manipulation strategy to engineer the phase transition in transition metal oxide.

In this work, epitaxial $VO_2$ crystal film with the thickness of about 20nm were prepared by an rf-plasma assisted oxide molecule beam epitaxy (OMBE) on $TiO_2$ (110) and (001) crystal substrates, respectively. Before the deposition, the $TiO_2$ substrates were ultrasonically cleaned in acetone and alcohol, and then rinsed in deionized water. Finally, the substrates were blown dry by nitrogen gas before introducing into the growth chamber with a base pressure better than $3 \times 10^{-9}$ Torr. The growth rate was calibrated by quartz crystal microbalance and the growth process was monitored by reflection high-energy electron diffraction.

The crystal structure of $VO_2$ epitaxial films were characterized by X-ray diffraction (XRD, Model D/Max 2550 V, Rigaku, Japan) with Cu Kα radiation (λ= 1.54178Å ). The atomic force microscopy (AFM, Dimension Icon from BRUKER) was measured in air at room temperature. The temperature dependent resistance tests fir the prepared samples were conducted by a home-made four-probe station with controllable temperature stage. $He^+$ Microscopy (HIM, from ZEISS Co.) was applied to inject $He^+$ in thin films with the dose of 1500 ions/$nm^2$ at 30kV. The infrared nano-imaging was recorded by scattering-type scanning near-field optical microscope (s-SNOM, from Neaspec GmbH Co.). The 10.632μm Mid-IR was used as the near-field illumination source laser, a homemade constant heater was linked to the s-SNOM.

The commercial software, COMSOL Multiphysics, was used to simulate the distribution of the phases and surface strain along the c axis of the $VO_2$ films. The 2-dimensional (2D) models of single layer $VO_2$ films with round (radius = 0.7 μm) and square (side length = 1.4 μm) $He^+$ injected areas were built based on a large pristine film (radius = 50 μm) with a round fixed constraint boundary. Meanwhile, the physical module of Structural Mechanics (Solid Mechanics) was applied to calculate the results.

To prepared high-oriented $VO_2$ film, $TiO_2$(001) and $TiO_2$ (110) crystal substrates were used for the film growth. Due to the close crystal lattice parameters for rutile $TiO_2$ (a=0.459 nm and c=0.296 nm) and rutile $VO_2$ (a=0.455, and c=0.285 nm), epitaxial $VO_2$ film can be well-grown by oxide molecular beam epitaxy technique (oxide MBE). Considering the growth facets as shown in Fig.1a and Fig.1d, the $c_R$ axis of the grown $VO_2$(001)/$TiO_2$(001) is out of plane, while the $c_R$ axis is in-plane for $TiO_2$ (110) substrate. The crystal structures were characterized by X-ray diffraction as shown in Fig.1b and 1e. The diffraction peaks at 62.74 and 65.30 °in Fig.1b are indexed to $TiO_2$ (002) and $VO_2$ (002) planes, respectively. No other peaks were shown by XRD analysis, indicating that the prepared film grown by oxide MBE is an (001)-oriented phase. In Fig.1e, the XRD patterns show sharp peaks at 27.44 related to the $TiO_2$ (110) and the peak at 27.68 is associated with the $VO_2$ $(110)_R$.

The temperature dependent resistance measurements for the prepared films were also conducted. In Fig 1c, sharp resistance change up to three-order of magnitude across the metal-insulator transition can be observed for VO$_2$ (001)/TiO$_2$ (001) film. While it should be noticed that the critical temperature was ~55°C and obviously lower than that of bulk VO$_2$ (~68°C). In Fig.1f, the R-T curve is also plotted for VO$_2$ (110)/TiO$_2$ (110) film, which shows quite broad hysteresis loop and the increased critical temperature. It was reasonable if considering the interfacial strain/stress from the substrate in epitaxial VO$_2$/TiO$_2$ system. The $c_R$ axis is compressed in VO$_2$(001)/TiO$_2$(001) film, while stretched in VO$_2$(110)/TiO$_2$(110), which resulting the different modulations of the phase transition temperature [13]. This different changes of the critical transition temperatures in these two films indicate the delicate correlation with the status of $c_R$-axis in VO$_2$ crystal.

After treated by He$^+$ nanobeam implantation with a square injection region, the VO$_2$ (001) /TiO$_2$ films were characterized by s-SNOM and AFM as showing in Fig.2. The scheme for the s-SNOM technique is shown in Fig.2a, which combines the advantages of phase sensitive of infrared light and the high spatial resolution of probe microscopy. By using the s-SNOM technique, it is clear to observe the phase transition of the VO$_2$ film sample after He$^+$ implantation with different naonbeam sizes in Fig.2b. It is suggested that for s-SNOM test, the detected signal S$_n$ is the light scattered by the tip demodulated at the *n*th harmonic of the tip tapping frequency [14,15]. Thus we choose $S_3$, the tip-scattered light demodulated at the third harmonic of the tapping frequency, to obtain the relatively useful signals considering the relationship between background signal and overall signal strength[15,16]. Fig.2c and 2d show the detailed s-SNOM images for VO$_2$ (001) sample after He$^+$ implantation with the beam size of 2μm × 2μm, while Fig.2e shows the AFM image for the same area. Since higher local optical conductivities yield stronger infrared near-field signal, the s-SNOM image with the bright square reveals that the He$^+$ injection area has been transferred from the original insulating to metallic phase. In addition, this phase transition is quite homogenous within the square area after He$^+$ injection. In fact, for the current He$^+$ radiated VO$_2$ (001) sample, we also conducted the temperature dependent resistance measurement (See supplementary FIG.1) for the He[+] radiated and untreated areas, respectively, which clearly showed the metallic feature and normal phase transition behavior, respectively. This observation is consistent with previous report[17], which demonstrated that Ar[+] implantation will induce the metallic $VO_2$ phase at room temperature. It is possible to highlight the degree of correlation between AFM topography and infrared near-field amplitude (S3) at the same coordinate system as shown in Fig.2f. The line scans of AFM topography (black curve) and the near-field $S_3$ amplitude (red curve) along the direction by the dotted arrow show almost the same profiles, indicating the unmistakable correlation between the infrared near-filed amplitude and topographic features for $VO_2$ (001)/ $TiO_2$ film.

While for $VO_2$ (110)/ $TiO_2$ film, the same He[+] beam treatment with a square region will induce quite different results as shown in Fig. 3. The s-SNOM images in Fig.3a and 3c reveal that the He[+] injected $VO_2$ (110) film can produce pronounced phase separations. Perpendicular to the $c_R$ axis direction, two bright stripes appears along two edges of the squared region. Since higher local optical conductivities yield stronger infrared near-field signal, it is suggested that there appears anisotropic metallic stripes caused by the He[+] flux treatment. The AFM mapping image in Fig.3b shows the clear protuberant square area after the He[+] implantation, which is similar as the observation for $VO_2$ (001) film in Fig.2e. The line scans of AFM topography (black curve) and the near-field S3 amplitude (red curve) along the direction by the dotted arrow for the ion-flux treated $VO_2$ (110)/ $TiO_2$ (110) film is also plotted in Fig. 3d. The height variation and the infrared near-filed amplitude is completely not consistent, which indicates different correlation between the two effects if comparing with the $VO_2$ (001)/ $TiO_2$ (001) film (Fig. 2f).

In fact, this different correlation depending on the growth orientation can be reproducible, even if changing the He[+] injection area from square to circle shape (See supplementary Fig.1 and Fig.2). It is interesting to see that the phase separations in the nanoscale range induced by the ion injection in $VO_2$(110)/$TiO_2$(110) films with the square shape are two parallel stripes, while two distinct arcs around the injection circle area shown in Fig.2S. In addition, it is also found that the metallic phase stripes are always perpendicular to the $c_R$ axis direction, showing the distinct anisotropic feature.

From the above results, it is deduced that the $c_R$ axis should play a key role in this nanoscale phase separation for the strained $VO_2$ film. In addition to the changing optical signal of the injected regions, the height variation can provide important hint about the underlying mechanisms that govern the MIT via strain. For the $VO_2(001)$ /$TiO_2$ (001) film, the $c_R$ axis is out-of-plane, the $He^+$ injection induced phase transition is homogenous and the anisotropic phase separation by ion injection is not observed. While for $VO_2(110)$ /$TiO_2$ (110) film, the $c_R$ axis is in-plane. Due to the $He^+$ insertion, the injected region will swell and form a big bump. Under this condition, the localized $VO_2$ crystal at the edges perpendicular to the $c_R$ direction will have a compressed $c_R$ value and lead to the metallic phase at the room temperature, which is clearly observed in the SNOM optical signal with anisotropic character in Fig.3 or Supplementary Fig.2. This anisotropic phase separation in nanoscale range can also be clearly observed in Fig.4. In fact, if we inject the $He^+$ flux onto $VO_2$ (110) film with the shape of number " 2 " or number "3" as shown in Fig.4a or 4c, the edges or arcs which are perpendicular to the $c_R$ axis, all show bright sections, indicating the metallic phase structures induced by the local $He^+$ flux treatment. The related 3D profile of the s-SNOM signal in Fig.4b or 4d further confirm the anisotropic features of the local phase separation.

To further explore the $c_R$ axis dependent anisotropic character of the phase separation for $VO_2(110)$ /$TiO_2$ (110) film, we just change the $He^+$ injection square region with a rotate angle (the square bottom edge and $c_R$ direction) from parallel to 45° and 60°, respectively. From Fig.5a~5c, it can be observed that for the initial stage, the two bright stripes are perpendicular to $c_R$ direction (Fig.5a). When rotating the square with 45°, the phase separation is distinct at the two vertex-angle along the $c_R$ direction. While if angle further changes to 60°, the two stripes also tilt to the same angle. These angle-dependent s-SNOM signals clearly confirm the anisotropic phase separation, which can be directly manipulated by adjusting the profile of the $He^+$ injection area.

The interfacial strain/stress simulation appears to provide a solid evidence for understanding this anisotropic phase separation in nanoscale. Since the strain distribution is quite closely associated with the phase separation on the film surface, the anisotropic strain distribution mapping as well as the local phase separation can be simulated at the same time (see supplementary Fig.3). Considering the He$^+$ injection square area, the simulations of the phase distribution mapping for VO$_2$/TiO$_2$(110) film with different orientations referred to the in-plane $c_R$ axis are conducted(see supplementary Fig.4), showing the strong $c_R$ axis dependence. Thus, comparing with the s-SNOM observations in Fig.5a~5c with a rotated square area, the simulations showing in Fig.5d~5f are quite consistent, confirming the anisotropic phase separation in nanoscale induced by the He$^+$ injection.

Considering the different symmetry of the He$^+$ injection shapes on VO$_2$/TiO$_2$(110) film (square or circle), the local the anisotropic strain distribution mapping are also simulated for both cases(see supplementary Fig.5). During the simulation, the anisotropic factors are well considered (see supplementary Note 1). For the circle shape (see supplementary Fig 6), the simulation results show that the increasing of the pressure can only raise the absolute value of the strain rather than affect the distribution of the strain, which directly reflected by the anisotropic phase separations in nanoscale size.

In summary, He$^+$ nanobeam injection into VO$_2$ film will induce homogenous metallic phase transition for VO$_2$ (001) film, while under the same condition, anisotropic phase separations are observed in nanoscale by s-SNOM in VO$_2$ (110) film, which show pronounced $c_R$-axis dependency. Combined with the AFM topography images, it is suggested that the He$^+$ implantation induced swelled region plays a key role in the phase separation in VO$_2$(110)/TiO$_2$ (110) epitaxial film. Theoretical simulations are quite consistent with the experimental observations, further confirming the phase separation induced by the compressed $c_R$ value in the local area, which can be directly manipulated by adjusting the He+ injection profile. The current studies not only reveal the $c_R$-axis dependent phase separation in VO$_2$ (110) film, but also provide an effective strategy to engineer phase transitions in nanoscale level for tuning the properties of correlated oxide system.

This work was partially supported by the National Natural Science Foundation of China (12074356)，the Open Research Fund of State Key Laboratory of Pulsed Power Laser Technology and the Youth Innovation Promotion Association CAS. This work was partially carried out at the USTC Center for Micro and Nanoscale Research and Fabrication. The authors also acknowledged the supports from the Anhui Laboratory of Advanced Photon Science and Technology.

## DATA AVAILABILITY

The data that support the findings of this study are available from the corresponding authors upon reasonable request.

# Figures and Captions

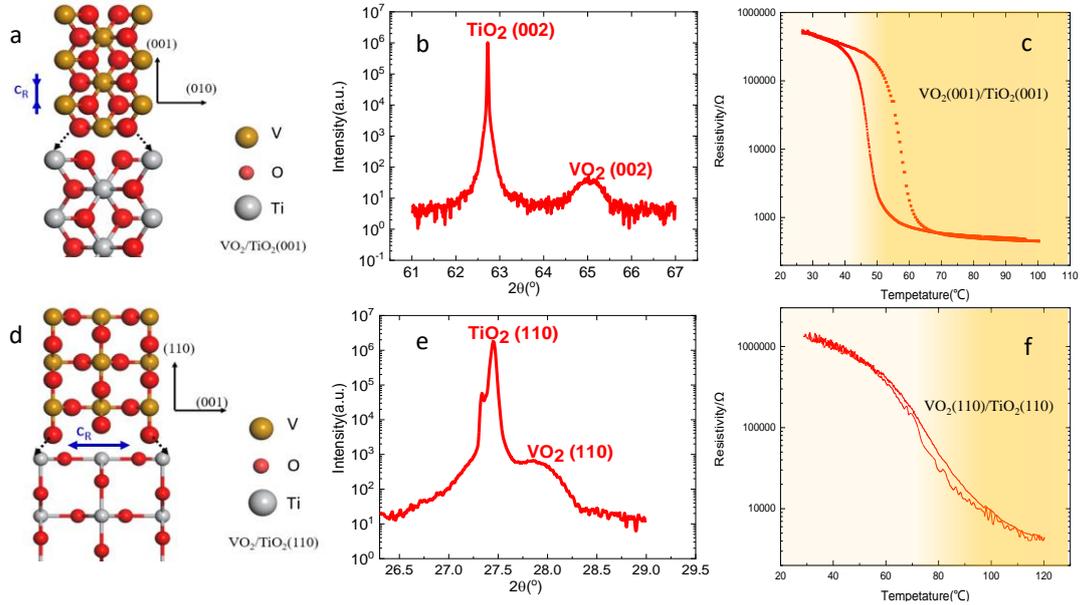

FIG. 1. Characterizations for the prepared VO$_2$ film. (a) Scheme for the epitaxial VO$_2$ (001) film grown on TiO$_2$ (001) crystal substrate; (b, c) The XRD and R-T curves for VO$_2$ (001) film. The phase transition temperature is lower than that of bulk due to the interfacial stress induced shrink $c_R$ axis; (d) Scheme for the epitaxial VO$_2$ (110) film grown on TiO$_2$ (110) crystal substrate; (e,f) The XRD and R-T curves for VO$_2$ (110) film. The phase transition temperature is higher than that of bulk due to the interfacial stress induced expand $c_R$ axis.

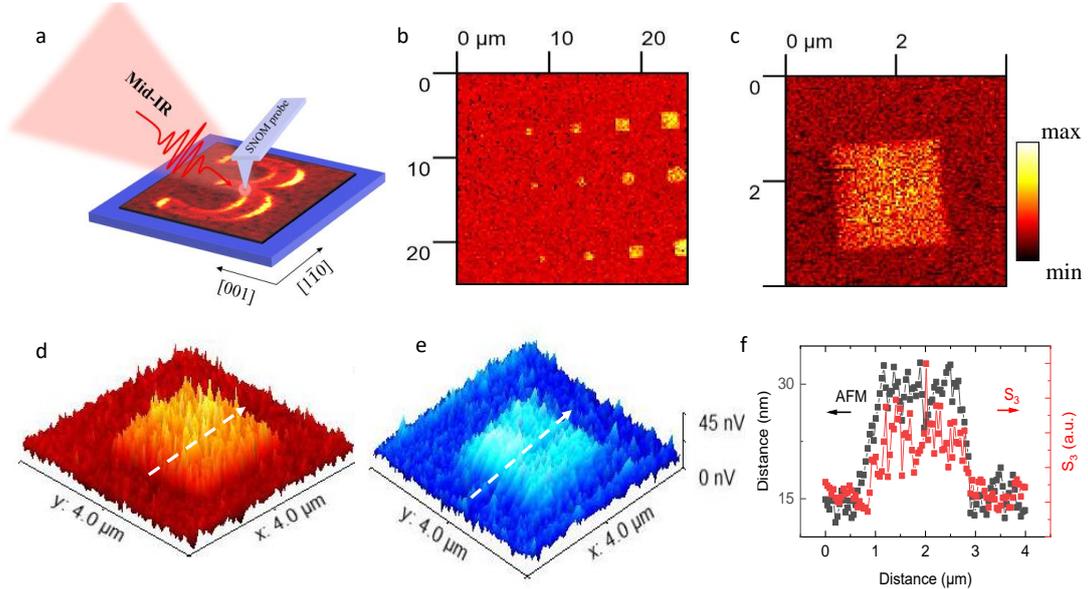

FIG. 2. The s-SNOM test for VO$_2$ (001)/TiO$_2$ film after He$^+$ implantation. (a) The scheme for the s-SNOM technique; (b) The s-SNOM image (scattering amplitude S$_3$ mapping) for the sample after He$^+$ implantation with different beam sizes. (c) The s-SNOM image for the sample after He$^+$ insertion with the size of 2μm × 2μm, showing a homogenous metallic state in the whole square area; (d) The 3D scattering amplitude S$_3$ image and (e) the 3D-AFM topography image for the film after He$^+$ insertion with the size of 2μm × 2μm; (f)The line scans of AFM topography (black curve) and the near-field S3 amplitude (red curve) along the dotted arrow lines in (d) and (e).

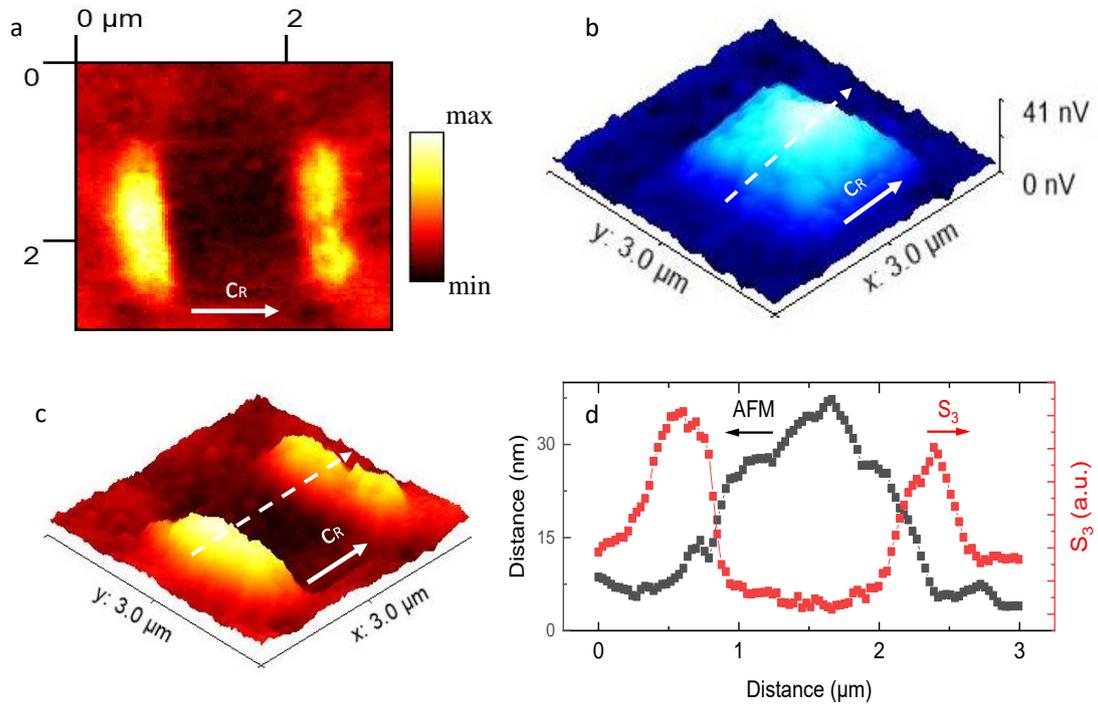

FIG. 3. The s-SNOM test for $VO_2(110)/TiO_2$ film after $He^+$ implantation. (a) The s-SNOM image (scattering amplitude $S_3$ mapping) for the sample after $He^+$ insertion, showing two stripe-like metallic phases perpendicular to $c_R$-axis; (b) The 3D-AFM topography image and (c) the 3D scattering amplitude $S_3$ image for the film after after $He^+$ insertion; (d)The line scans of AFM topography (black curve) and the near-field S3 amplitude (red curve) along the dotted arrow lines in (b) and (c).

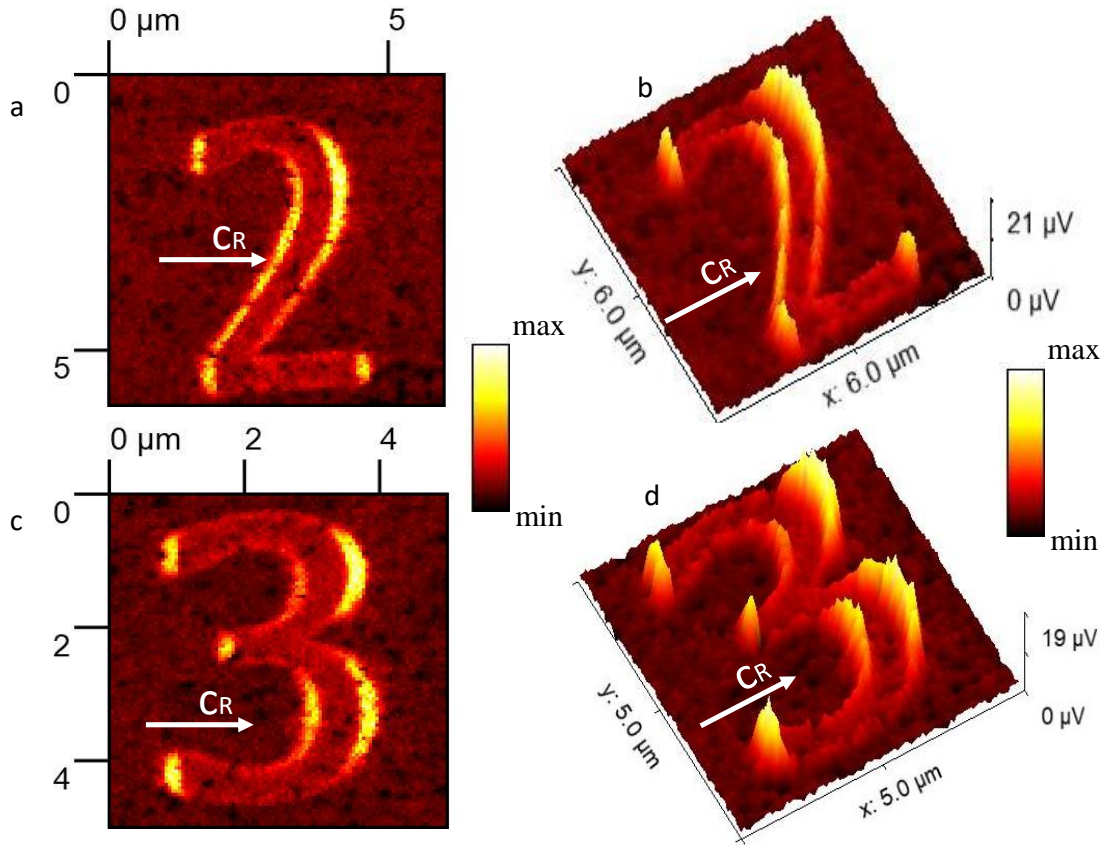

FIG.4. The s-SNOM image (scattering amplitude S3 mapping) for the VO$_2$ grown TiO$_2$(110) after He$^+$ insertion with circle area with the shape of number "2" (a) and number "3" (c), the (b) and (d) representing the 3D scattering amplitude S3 image of (a) and (c), respectively.

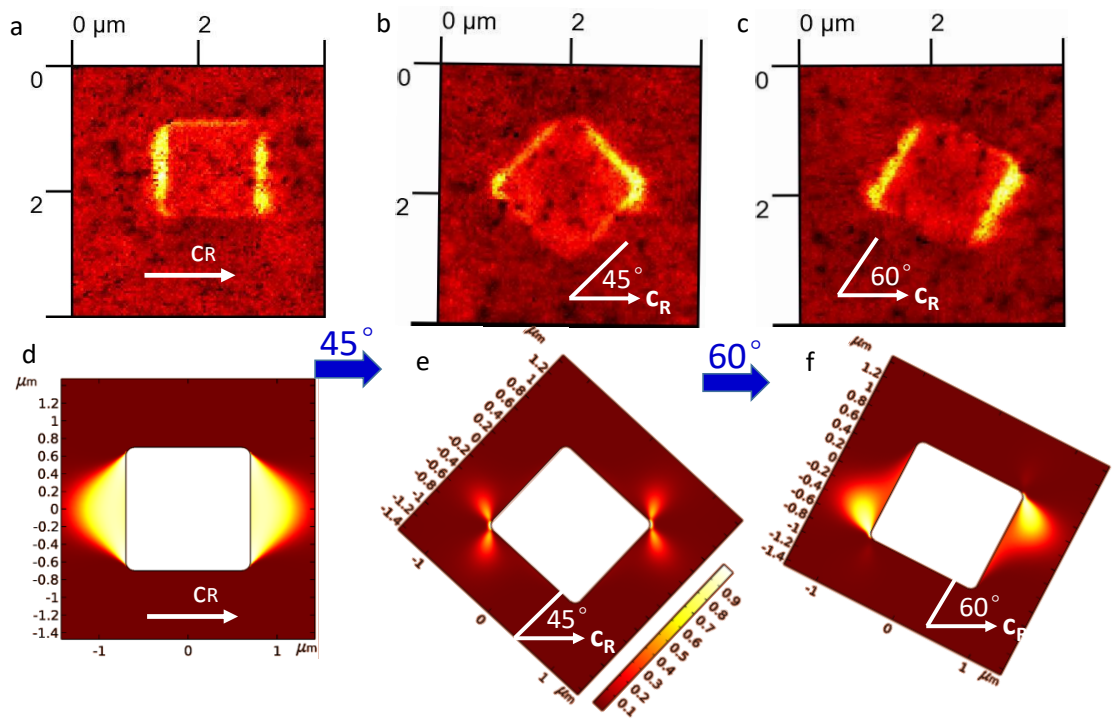

FIG.5. Experimental and Simulations for the anisotropic phase separation. (a~c) The s-SNOM images (scattering amplitude $S_3$ mapping) for the sample after He$^+$ insertion by changing the angle between the square bottom edge and $c_R$ direction; (d~f) The simulations for the anisotropic phase separation depending on $c_R$ axis direction, consistent with the s-SNOM observations for the anisotropic phase separation.